\documentclass[journal=jacsat,manuscript=article]{achemso}
\setkeys{acs}{maxauthors = 5}
\setkeys{acs}{etalmode = truncate}

\usepackage{graphicx}
\usepackage{dcolumn}
\usepackage{bm}
\usepackage{multirow}
\usepackage{multicol}
\usepackage[ruled]{algorithm2e}
\usepackage{subfigure}
\usepackage{bbding}

\usepackage{graphicx}
\usepackage{dcolumn}
\usepackage{bm}
\usepackage{mathrsfs}
\usepackage{amsmath}
\usepackage{float}
\usepackage{enumitem}
\usepackage{subfigure}
\usepackage{tabulary}
\usepackage{soul}
\usepackage{amssymb}

\usepackage{color}
\usepackage{xcolor}
\usepackage{listings}

\definecolor{dblue}{rgb}{0,0,0.7}
\definecolor{lgray}{rgb}{.4,.4,.4}

\newcommand{\system}{\textit{Q$^2$Chemistry}}

\SectionNumbersOn

\title{Q$^2$Chemistry: A quantum computation platform \\ for quantum chemistry}

\author{Yi Fan}
\affiliation{Hefei National Research Center for Physical Sciences at the Microscale, University of Science and Technology of China, Hefei, Anhui 230026, China}
\author{Jie Liu}
\affiliation{Hefei National Research Center for Physical Sciences at the Microscale, University of Science and Technology of China, Hefei, Anhui 230026, China}
\author{Xiongzhi Zeng}
\affiliation{Hefei National Research Center for Physical Sciences at the Microscale, University of Science and Technology of China, Hefei, Anhui 230026, China}
\author{Zhiqian Xu}
\affiliation{Hefei National Research Center for Physical Sciences at the Microscale, University of Science and Technology of China, Hefei, Anhui 230026, China}
\author{Honghui Shang}
\affiliation{Hefei National Research Center for Physical Sciences at the Microscale, University of Science and Technology of China, Hefei, Anhui 230026, China}
\author{Zhenyu Li}
\email{zyli@ustc.edu.cn}
\affiliation{Hefei National Research Center for Physical Sciences at the Microscale, University of Science and Technology of China, Hefei, Anhui 230026, China}
\author{Jinlong Yang}
\affiliation{Hefei National Research Center for Physical Sciences at the Microscale, University of Science and Technology of China, Hefei, Anhui 230026, China}

\begin{document}


\date{\today}

\begin{abstract}
Quantum computer provides new opportunities for quantum chemistry. In this article, we present a versatile, extensible, and efficient software package, named {\system}, for developing quantum algorithms and quantum inspired classical algorithms in the field of quantum chemistry. In {\system}, wave function and Hamiltonian can be conveniently mapped into the qubit space, then quantum circuits can be generated according to a specific quantum algorithm already implemented in the package or newly developed by the users. The generated circuits can be dispatched to either a physical quantum computer, if available, or to the internal virtual quantum computer realized by simulating quantum circuit on classical supercomputers. As demonstrated by our benchmark simulations with up to 72 qubit, {\system} achieves excellent performance in simulating medium scale quantum circuits. Application of {\system} to simulate molecules and periodic systems are given with performance analysis. 
\end{abstract}

\maketitle

\section{Introduction}
 As the application of accurate classical methods are severely limited by the fast growing computational cost, quantum computation provides a promising pathway to solve the quantum chemistry problems\cite{Preskill2018quantumcomputingin, McAEndAsp20, YungCas2014}. By encoding wave functions into the \textit{Hilbert} space of qubits, the Schrodinger equation for molecular systems can be solved on a quantum computer. In recent years, various quantum algorithms, such as quantum phase estimation (QPE) and variational quantum eigensolver (VQE),\cite{Tilly_VQE_2021, Cerezo2021, Magann_VQA_2021, Fedorov_VQERev_2022, BraKit02, McAEndAsp20, CaoRomOls19, Pre18, GeoAshNor14, AspDutLov05, Wang08, PerMcCSha14, HemMaiRom18, NamChen20, SheZhaZha17, MalBabKiv16, KanMezTem17, ColRamDah18, McCRomBab16, LanWhi10, RomBabMcC18, vqe-excited-vqd, Mcclean17qse, YungCas2014}  are developed for quantum chemistry.  There are some important issues should be studied for these algorithms. For example, the basis set used to construct the electronic Hamiltonian and the fermion-to-qubit encoding technique may lead to significant gate complexity and measurement overhead in QPE\cite{Liu_BASIS_REV_2022}. The optimization of typical wave function ansatzes using VQE can suffer from the "barren plateaus"\cite{McClean_BARREN_2018, John_BARREN_2022} or local minimum traps\cite{Anschuetz_MINIMA_TRAP_2022}. A quantum computation platform which can provided extensive functionalities and step-by-step cross verification is therefore desirable for designing novel quantum algorithms for chemistry applications.

Noisy intermediate quantum (NISQ) devices  have been used to demonstrate the possility of studying the ground and excited states of molecular systems  using currently available algorithms.\cite{Kandala2017, PerMcCSha14, HemMaiRom18, NamChen20, SheZhaZha17, MalBabKiv16, KanMezTem17, ColRamDah18, Google_HF_2020, Huggins_QMC_EXPERI_2022} However, such experimental demonstrations are limited to tiny molecules with an artificially small basis set. This is because NISQ experiments are limited by the available quantum resources and the error associated with each quantum gate. Therefore, low gate fidelity, short coherence time, and insufficient qubit resource prohibit a systematic study of quantum chemistry oreinted quantum algorithms on NISQ devices. The largest quantum computation experiment for chemistry up to date uses 16 qubits with 160 two-qubit gates\cite{Huggins_QMC_EXPERI_2022}, while a simple VQE circuit of the commonly used unitary coupled-cluster (UCC)\cite{Kut82, BarKucNog89, TauBar06} ansatz for a small molecule has $10^7$ CNOT gates (Table~\ref{tab:resource_est}), which is far beyond the capability of NISQ devices. Therefore, it is important to have the capability to simulate quantum circuit on a classical computer at this stage. Even beyond the NISQ era, such a capability can help us to develop quantum inspired classical algorithms.  

\begin{table}
\caption{Computational resources required to perform VQE simulations for a number of molecules using the unitary coupled-cluster ansatz truncated up to double excitations (UCCSD) and the minimum basis set STO-3G. UCCGSD means that generalized excitation operators (not distinguishing occupied and virtual orbitals) are used. }
\centering
\begin{tabular}{cccccc}
\hline \hline
{Molecule}  & {Qubits}  & \multicolumn{2}{c}{Parameters} & \multicolumn{2}{c}{CNOT gates} \\
                               &                            & UCCSD & UCCGSD & UCCSD & UCCGSD \\
\hline 
H$_2$ & 4 & 2 & 5 & 64 & 144 \\
LiH & 8 & 14 & 72 & 1632 & 10720 \\
H$_2$O & 14 & 90 & 630 & 26272 & $2.1\times 10^5$ \\
NH$_3$ & 16 & 135 & 1064 & 46480 & $4.4\times 10^5$ \\
CH$_4$ & 18 & 230 & 1692 & 95200 & $8.2\times 10^5$ \\
C$_2$H$_4$ & 28 & 1224 & 9737 & $8.6\times 10^5$ & $7.9\times10^6$ \\
C$_3$H$_6$ & 42 & 5994 & 48930 & $6.6\times 10^6$ & $6.3\times 10^7$ \\
\hline \hline
\end{tabular}
\label{tab:resource_est}
\end{table}

Several quantum computation packages have been reported, for example, C++ based ProjectQ\cite{projectq} and Qiskit\cite{Qiskit}, GPU-enabled Qulacs\cite{qulacs}, and the differentiable simulator Yao\cite{yao} implemented in Julia. Most of these codes are developed as stand-along quantum circuit simulators or compilers, which have no interfaces for quantum chemistry applications. Some of the packages such as Qiskit or PennyLane\cite{pennylane} provide modules linked to external ab-initio chemistry codes. However, the functionalities are limited, for example, with support for periodic boundary condition which plays an important role in modeling materials. At the same time, the performance of simulating quantum circuit is not very satisfactory in these packages. Most of these packages only implement the brute-force simulating method which leads to an exponential computational cost. Tensor-based methods have been implemented in packages such as Qiskit and PennyLane, however, without an efficient distributed parallelization algorithm. As a reference, the largest simulated quantum circuit up to date contains 28 qubits, which is used to study the ground state of ethene molecule using VQE\cite{cao_SYMM_2022}.

Based on the above considerations, we develop a versatile and extensible quantum computation platform for quantum chemistry, named {\system} (pronounced as "Q square chemsitry") to highlight the two quantum dimensions (the systems to be studied and the tools used to study them).  {\system} adopt a modular design and a mixed-language programming model to achieve versatility together with high performance, where Python is used as its application program interface (API) while C++ and Julia\cite{Bezanson_JULIA_2017} are used for computational intensive tasks such as quantum circuit simulators. {\system} provides interfaces to quantum chemistry packages such as PySCF\cite{pyscf} to generate required parameters for the qubit Hamiltonian. The quantum algorithms for solving the eigenstates of the electronic Hamiltonian are programmable at a high level of abstraction by adopting internal templates and/or defining custom circuit generation procedures. Some popular VQE and post-VQE algorithms have already been implemented for ground and excited states of molecular and periodic systems. {\system} provides native modules to run the circuit generated by a specific quantum algorithm, either via various high performance classical circuit simulation algorithms for simulating the circuit on a classical computer, or via extensible interfaces reserved for the upcoming actual quantum processors. An efficient matrix product state (MPS)\cite{Orus_MPS_PEPES_2014, Schollwock_DMRG_MPS_2011} based circuit simulation engine is implemented to perform quantum simulations for systems with 50$\sim$100 qubits. 

The remaining part ot this article is organizing as the following. After introducing the framework of {\system}  in Section~\ref{sec:design}, we present functionalities for handling ab-initio chemistry quantities in Section~\ref{sec:qc}. In Section~\ref{sec:qcirc} we explain the implementation of the circuit simulation on a classical computer or running on the upcoming quantum hardwares. Section~\ref{sec:qalgo} gives a general introduction to the natively implemented VQE-based algorithms. Finally, in Section~\ref{sec:app}, we provide some examples to demonstrate the power of {\system} in chemistry applications, and provide the road map for further extensions including the support for more quantum algorithms, classical circuit simulation backends, and circuit optimization algorithms.

\section{The framework of {\system}}
\label{sec:design}
\begin{figure*}
    \centering
    \subfigure[]
    {
        \includegraphics[width=0.61\textwidth]{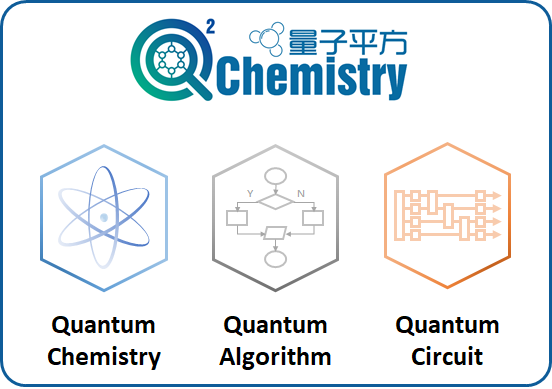}
    }
    \subfigure[]
    {
        \includegraphics[width=0.1703125\textwidth]{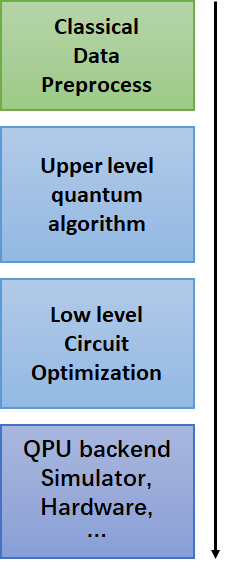}
    }
    \caption{(a) The framework of {\system}. (b) A typical workflow of solving a chemical problem using a quantum algorithm.}
    \label{fig:framework}
\end{figure*}

As shown in Figure~\ref{fig:framework}a, {\system} contains three modules. (1) \texttt{q2chem.qchem}: the quantum chemistry module which defines quantum chemistry problems in the qubit space typically with the help of external classical quantum chemistry packages; (2) \texttt{q2chem.qcirc}: the quantum circuit module which provides quantum circuit related functionalities, including circuit construction, visualization, optimization, and execution on a virtual or real quantum computer; (3) \texttt{q2chem.qalgo}: the quantum algorithm module which includes native quantum algorithms and also provides tools to implement new algorithms to solve chemistry problems.

High-level modules in {\system} are implemented using the scripting language Python and provide base implementations for classes in submodules. Benefiting from the modular design and Python's extensibility, extending existing submodules are easily achieved by constructing derived classes and implementing only a small number of virtual functions, which requires no modifications to higher level modules. Core functions in the backends are implemented using programming languages including C++ and Julia. These functions are integrated into the Python interfaces using just-in-time (JIT) technology, which compiles the code at run time to provide machine-specific optimizations and deliver outstanding performance. Most of the low-level data structures accessible from Python are stored using NumPy's\cite{numpy} \texttt{ndarray}. Therefore, auxiliary operations, such as exact diagonalization for the qubit Hamiltonian, can be realized by using NumPy-compatible packages such as SciPy\cite{scipy} or PyTorch\cite{pytorch}.

Using such a modular framework, a general workflow to solve a quantum chemistry problem on a quantum computer is briefly illustrated in Figure~\ref{fig:framework}b. (1) Collect classical data such as electron integrals and mean-field orbital coefficients to generate Hamiltonian and an initial quantum state. (2) Choose or develop a suitable quantum algorithm and generate corresponding quantum circuits, which may depend on the Hamiltonian such as in the QPE algorithm or be system independent such as in some VQE algorithms. (3) Perform lower level circuit optimizations, including eliminating redundant quantum gates to reduce circuit depth or reconstructing the circuit to fit a specific quantum processor.(4) Execute the quantum circuits and perform some measurements to extract necessary information, using either a virtual or real quantum computer.

Currently, {\system} can be routinely used for performing VQE simulations on a classical computer, which is powered by Hamiltonian generation for molecular and periodic systems, hardware-efficient and UCC-based ansatzes for parametric circuit construction, and a scalable noise-free tensor network backend for classical simulations of large quantum circuits.

\section{The quantum chemistry module}\label{sec:qc}
The \texttt{q2chem.qchem} module mainly handles system Hamiltonian and wave function mapping. The second-quantized Hamiltonian is constructed from classically calculated quantities such as molecular orbital coefficients. The wave function mapping determines how the orbitals of simulated wave function are mapped onto qubits therefore influences measurement strategy.

\subsection{Hamiltonian}\label{sec:qc:ham}
For a Hamiltonian which is expressed as the linear combination of Pauli strings
\begin{equation}
    \hat{H} = \sum_{i}{c_{i} \hat{P}_{i}}
    \label{eq:qubit-ham}
\end{equation}
where $\hat{P}_{i}$ is product of Pauli operators $\{\hat{\sigma}_{x}, \hat{\sigma}_{y}, \hat{\sigma}_{z}\}^{\otimes}$, the expectation value $E$ can be obtained through quantum measurement techniques such as the Hadamard test as
\begin{equation}
    \begin{aligned}
        E &= \langle \Psi | \hat{H} | \Psi \rangle \\
        &= \langle \Psi | \sum_{i}{c_{i} \hat{P}_{i}} | \Psi \rangle \\
        &= \sum_{i}{c_{i} \langle \Psi | \hat{P}_{i} | \Psi \rangle}.
    \end{aligned}
    \label{eq:qubit-ham-detail}
\end{equation}
Given the Hartree-Fock orbitals, the second-quantized electronic Hamiltonian can be written as
\begin{equation}
    \hat{H} = \sum_{p, q}{h_{pq} \hat{T}^{p}_{q}}  + \sum_{\substack{p, q\\r, s}}{g^{pq}_{rs}\hat{T}^{pq}_{rs}},
    \label{eq:ferm-ham}
\end{equation}
with
\begin{equation}
    \begin{aligned}
        \hat{T}^{p}_{q} &= a^{\dagger}_{p} a_{q} \\
        \hat{T}^{pq}_{rs} &= a^{\dagger}_{p} a^{\dagger}_{q} a_{r} a_{s}.
    \end{aligned}
\end{equation}
$h_{pq}$ and $g^{pq}_{rs}$ are one- and two-electron integrals. In order to convert the second-quantized Hamiltonian Equation~\ref{eq:ferm-ham} into the qubit form as given in Equation~\ref{eq:qubit-ham}, a fermion-to-qubit mapping such as the Jordan-Wigner or Bravyi-Kitaev mapping is required\cite{JordanWigner28, SeeRich12,Tran18}.

In {\system}, an interface with the PySCF package is provided to calculate one- and two-electron integrals. for molecular and periodic systems\cite{Liu_PBC_2020, FAN_PBC_2021}. Orbital optimization is also supported by linking to the \texttt{pyscf.lo} module or using a custom cost function. A unitary matrix is then obtained to transform the integrals as:
\begin{equation}
    \begin{aligned}
        \tilde{h}_{\tilde{p} \tilde{q}} &= \sum_{p,q}{U_{p \tilde{p}} h_{pq} U_{q \tilde{q}}} \\
        \tilde{g}^{\tilde{p} \tilde{q}}_{ \tilde{r} \tilde{s}} &= \sum_{\substack{p, q\\r, s}} {g^{pq}_{rs} U_{p \tilde{p} } U_{q \tilde{q} } U_{r \tilde{r}} U_{s \tilde{s} }}.
    \end{aligned}
\end{equation}
This step is carried out efficiently by calling the optimized tensor contraction package \texttt{opt\_einsum}\cite{opteinsum}. For fermion-to-qubit mapping, {\system} implements an efficient Jordan-Wigner transformation written in pure Julia. Other methods such as Bravyi-Kitaev are currently provided through the interface of OpenFermion\cite{openfermion}. The qubit Hamiltonian can then be used to construct quantum gates for measurement in quantum algorithms such as VQE or QPE.

\subsection{Wave function}\label{sec:qc:wfn}
To represent a many-electron quantum state on a quantum computer, the most commonly used strategy is to map its molecular orbitals onto qubits of which can span the Fock space. Such a straightforward orbital-to-qubit approach can be written as
\begin{equation}
    \begin{aligned}
        | \Psi_{HF} \rangle &= |i_{0} i_{1} \cdots \rangle \\
        | \Psi_{CI} \rangle &= \sum_{i_{0},i_{1}, \cdots} { c_{i_{0} i_{1}\cdots} |i_{0} i_{1} \cdots \rangle },
    \end{aligned}
    \label{eq:jw-wfn-map}
\end{equation}
where $i_{j} \in \{0, 1\}$ represents both the occupation of orbitals and the quantum state $|0\rangle$ or $|1\rangle$ of the corresponding qubit. Equation~\ref{eq:jw-wfn-map} actually describes the quantum state corresponding to the eigenstates of a qubit Hamiltonian which is obtained from the Jordan-Wigner transformation. If other fermion-to-qubit mapping algorithms such as Bravyi-Kitaev are used for the Hamiltonian, such a correspondence does not necessarily exist. With a quantum state mapped to qubits, {\system} provides Hadamard test to evaluate the expectation value of the quantum state with respect to an operator. 

In the simple orbital-to-qubit mapping, the number of qubits required to simulate the wave function has a linear dependence on the number of basis functions, which prohibit the use of a large basis set on NISQ devices. {\system} provides another strategy which maps a classical tensor network (TN) state onto quantum circuits\cite{Liu_FEW_QUBIT_2019, Haghshenas_QCTN_2022}. In this way, the qubits determines the classical bond dimension of the tensor network. For chemical systems which contain weak electron correlations or have a special symmetry, such a TN-based strategy provides a possible solution to effectively reduce the number of qubits at the expense of performing more measurements. Generally, different wave function mapping strategy leads to distinctive structures of the quantum circuits, therefore may bring special restrictions to subsequent quantum algorithms and measurement protocols. The \texttt{q2chem.qchem} module passes the mapping strategy to quantum algorithms in \texttt{q2chem.qalgo} to generate an abstract circuit class for initialization. The abstract circuits are then extended according to the adopted quantum algorithm and instantiated to a common readable quantum circuit that can be used in the \texttt{q2chem.qcirc}.

\section{The quantum circuit module}\label{sec:qcirc}
The \texttt{q2chem.qcirc} module provides quantum circuit execution functionalities. It mainly contains two parts: a interface reserved for quantum computer manufacturers and a classical simulator which simulates state evolution determined by quantum circuits on a classical computer.

\subsection{Interfaces to quantum devices}\label{sec:qcirc:hardware-and-custom}

\begin{lstlisting}[float=*t,caption={Interface to quantum devices in the \texttt{qcircuit} module.},xleftmargin=21pt,framexleftmargin=17pt, label=code:simulator]
class _BaseQPU(object):
    def __init__(self, options, ...):
        self.quantum_state = ...
        ...
    def evolve_circuit(self, ...):
        ...

class _BaseHardware(_BaseQPU):
    def evolve_circuit(self, ...):
        circuit_asm = self._compile_circuit(...)
        measurement_result = self._quantum_hardware_evolution(...)
        ...
    def _compile_circuit(self, ...):
        # Implement for specific hardware.
    def _quantum_hardware_evolution(self, ...):
        # Implement for specific hardware.

class _BaseSimulator(_BaseQPU):
    def evolve_circuit(self, ...):
        ...
        for op in quantum_circuit:
            if isinstance(op, QuantumGate):
                self._quantum_gate_evolution(...)
            elif isinstance(op, MeasurementOp):
                self._measure_qubit(...)
        ...
    def _quantum_gate_evolution(self, ...):
        # Implement in custom simulators.
    def _measure_qubit(self, ...):
        # Implement in custom simulators.
    
\end{lstlisting}

Functionalities in \texttt{q2chem.qcirc} are inherited from the \texttt{\_Base} class. High-level Python APIs enable efficient extensions of {\system}, which can be used to interface with different quantum devices operation systems. Currently, there are multiple competing technical routes of quantum computer, e.g., the photonic qubits have long coherence time while superconducting platforms have good scalability. In order to efficiently and accurately study quantum chemical problems, various quantum hardwares may be required depending on the characteristics of the quantum algorithm. Therefore, we provide a flexible interface for connecting to different quantum hardware platforms as illustrated in Code Example~\ref{code:simulator}. In a general \texttt{\_BaseHardware}, the circuit generated by {\system} is firstly represented by quantum assembly language such as OpenQASM via \texttt{\_compile\_circuit()}. This step generally should include circuit optimization and reconstruction to fit the architecture of specific hardwares. Then the sequence of gates are converted into signals or pulses and sent to the quantum devices through \texttt{\_quantum\_hardware\_evolution()} using a compiler provided by the vendor of the hardware system. Finally, results such as measurement statistics or bit strings are collected and post-processed in {\system} to obtain required quantities.

\subsection{Classical quantum circuit simulator}

\begin{table*}
\caption{Backends for circuit simulations in present open-source quantum computation softwares. External indicates that one or more third-party packages are required to enable the functionality. SA stands for single-amplitude simulation.}
\centering
\begin{tabular}{cccccccc}
\hline \hline
\multirow{2}{1.5cm}{Software}  & \multicolumn{3}{c}{Backend} & \multicolumn{2}{c}{Parallism} & \multirow{2}{1.5cm}{GPU} & \multirow{2}{1.8cm}{Efficient Gradients} \\
                               & SV & DM & TN                & Threaded & Distributed       &     \\
\hline 
ProjectQ\cite{projectq}           & $\checkmark$ & Ongoing      & $\times$    & $\checkmark$ & SV           & $\times$     & $\times$      \\
HiQ\cite{hiq-projectq}            & $\checkmark$ & Ongoing      & SA          & $\checkmark$ & SV           & Ongoing      & $\times$      \\
MindQuantum\cite{mindquantum}     & $\checkmark$ & $\checkmark$ & $\times$    & $\checkmark$ & $\times$     & $\checkmark$ & $\checkmark$  \\
Qulacs\cite{qulacs}               & $\checkmark$ & $\checkmark$ & $\times$    & $\checkmark$ & $\times$     & $\checkmark$ & $\times$      \\
Yao\cite{yao}                     & $\checkmark$ & $\checkmark$ & Ongoing     & $\checkmark$ & $\times$     & $\checkmark$ & $\checkmark$  \\
Qiskit\cite{Qiskit}               & $\checkmark$ & $\checkmark$ & MPS         & $\checkmark$ & SV           & $\checkmark$ & $\times$      \\
Cirq\cite{cirq}                   & $\checkmark$ & $\checkmark$ & External    & $\checkmark$ & $\times$     & $\checkmark$ & $\times$      \\
PaddleQuantum\cite{Paddlequantum} & $\checkmark$ & $\checkmark$ & $\times$    & $\checkmark$ & $\times$     & $\checkmark$ & $\checkmark$  \\
PennyLane\cite{pennylane}         & $\checkmark$ & $\checkmark$ & MPS         & $\checkmark$ & $\times$     & External     & $\times$      \\
QuEST\cite{Jones_QUEST_2019}      & $\checkmark$ & $\checkmark$ & $\times$    & $\checkmark$ & SV, DM       & $\checkmark$ & $\times$      \\
{\system}                           & $\checkmark$ & $\checkmark$ & MPS         & $\checkmark$ & $\checkmark$ & $\checkmark$ & $\checkmark$  \\

\hline \hline
\end{tabular}
\label{tab:quantum_soft}
\end{table*}

{\system} implements several simulators to run quantum circuits on a classical computer. The quantum state of multiple qubits can be expanded using a certain basis set
\begin{equation}
    \label{eq:sv}
	\left| \Psi \right \rangle = \sum_{i_1 i_2 \ldots i_N} {c_{i_1 i_2 \ldots i_N} |i_1 i_2 \ldots i_N \rangle},
\end{equation}
where $N$ is the number of qubits, and $|i_1 i_2 \ldots i_N\rangle$ is the basis state. The coefficients $c_{i_1 i_2 \ldots i_N}$ is an $N$-dimensional tensor which contains $2^N$ amplitudes. Equation~\ref{eq:sv} is similar to the correlated wave function in quantum chemistry represented by a configuration interaction (CI) expansion. 

Currently, there are two strategies to simulate the evolution of quantum state on a classical computer. One is brute-force simulation which uses a ($2^{N}\times 1$) state vector (SV) or a ($2^{N} \times 2^{N}$) density matrix (DM) to represent the quantum state. The other is tensor network methods which approximate the quantum state by a set of low-rank tensors. The brute-force methods are supported by most of the existing packages (Table~\ref{tab:quantum_soft}).  {\system} also supports the state vector and density matrix methods. These brute-force methods implemented in the \texttt{q2chem.qcirc} module uses compressed sparse row (CSR) format for sparse matrix storage and OpenMP for multi-threaded calculation. An OpenACC-enabled C++ code is also implemented and can be selected at compile time to utilize GPU, which can provide $2\sim10$ times speed-up over the multi-threaded CPU version. 

A typical drawback of the brute-force simulation is that the memory usage and computational complexity both scales as $\mathcal{O}(c^{N})$, where $c$ is a constant between 2 to 4 and $N$ is the number of qubits. This exponential scaling prevents quantum simulations for larger molecules with more than $\sim$30 qubits. Recently, matrix product states (MPS) and projected entangled pair states (PEPS) are used to simulate large scale random quantum circuits\cite{Guo_QsimPEPS_2019, Guo_QsimTN_2021, Liu_QsimTN_SW_2021, McCaskey_QsimMPS_2018}. Since the tensor contraction pattern is mostly fixed and does not have the NP-hard path optimization problems\cite{Chung_TN_PATH_1997, Pfeifer_TN_PATH_2014}, the one-dimensional MPS is preferable as a high performance tensor network backend for a general quantum circuit simulator.

The MPS ansatz factorizes the rank-$N$ coefficients into lower rank tensors which can be written as
	\begin{equation}
		\label{eq:mps-rep}
		c_{i_1 i_2 \ldots i_N} = \sum_{u_0\ldots u_N} {{^1T^{i_1}_{u_0 u_1}} {^2T^{i_2}_{u_1 u_2}} \ldots {^NT^{i_N}_{u_{N-1} u_{N}}} }
\end{equation}
where ${^k T^{i_k}_{u_{k-1} u_{k}}}$ is a rank-$3$ tensor with $i_k$ called the \textit{physical} index and $u_k$ the \textit{auxiliary} index. The maximum size of the auxiliary indices are defined as the \textit{bond dimension} of the MPS, which is denoted as $D=\max_{0\le k \le N}{\{u_k\}}$. Algorithms based on MPS generally have a complexity of $\mathcal{O}(ND^3)$. If the bond dimension $D$ is allowed to grow exponentially, the MPS can exactly represent any quantum state using Equation~\ref{eq:mps-rep}.

\begin{figure}
    \centering
    \subfigure[]
    {
        \includegraphics[width=0.092\textwidth]{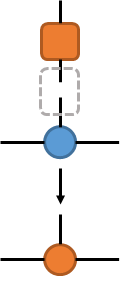}
    }
    \subfigure[]
    {
        \includegraphics[width=0.136\textwidth]{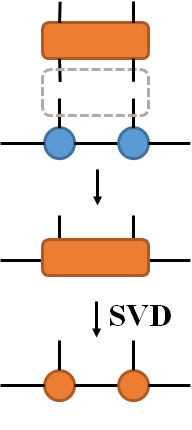}
    }
    \subfigure[]
    {
        \includegraphics[width=0.3\textwidth]{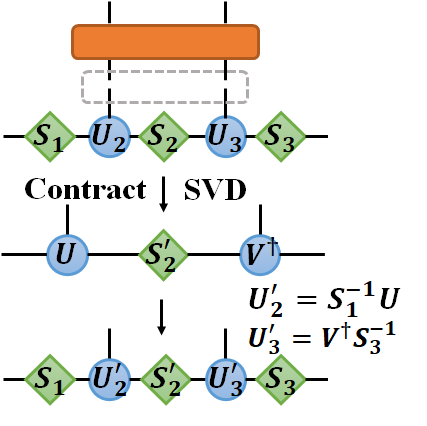}
    }
    \caption{(a) Applying a single qubit gate on the MPS quantum state is simulated by simply a local contraction. (b) Applying a two-qubit gate on neighbouring qubits generally has 2 steps: 1) Reshape the two-qubit gate into a 4-dimensional tensor and contract with the qubits to form a two-qubit tensor; 2) perform a singular value decomposition to restore the two-qubit tensor back to the MPS formulation. Post-processing is usually required to maintain normalization or canonicalization of MPS tensors; (c) Auxiliary matrices which contains truncated and normalized singular values are used to achieve normalization of the quantum state.}
    \label{fig:mps-algo}
\end{figure}

The \texttt{q2chem.qcirc} module implements the MPS simulation algorithm on the top of the machine learning framework PyTorch\cite{pytorch} based on the algorithm proposed by Guifr\'e \textit{et al.}\cite{Vidal_MPS_Qsim_2003}. A demonstrative procedure for simulating quantum circuit using MPS algorithm is presented in Figure~\ref{fig:mps-algo}. Different from the common classical MPS-based methods such as density matrix renormalization group (DMRG)\cite{Schollwock_DMRG_MPS_2011}, a set of auxiliary matrices are inserted between the rank-3 tensors and stored to maintain a normalized quantum state after the truncated SVD.

Classical quantum simulator backends can also be extended within the framework of {\system} in a similar style as the interfaces for hardwares. It requires only a minimum effort of implementing functions \texttt{\_quantum\_gate\_evolution()} and \texttt{\_measure\_qubit()} for a derived class of \texttt{\_BaseSimulator} and, if necessary, a custom data structure to store the quantum state. No modifications to upper level modules such as expectation value evaluation or higher level quantum algorithms is required. In Section~\ref{sec:app:result}, we show the application of {\system} interfaced with an external MPS simulator QuantumSpins\cite{Guo_GITHUB_2020, Guo_QsimTN_2021} which efficiently simulated a 40-qubit molecule with a high accuracy.

\subsection{Reverse-mode differentiation}

In many quantum algorithms, a parametric quantum circuit is constructed and optimization of the circuit parameters is carried out iteratively. On a quantum computer, the gradients of energy or another target function with respect to circuit parameters can be calculated through the parameter-shift rule or finite difference steps, which will introduce an additional complexity factor of $\mathcal{O}({N_{p}})$, where ${N_{p}}$ is the number of parameters.
Nevertheless, using a classical quantum circuit simulator the gradients can be evaluated efficiently with an approximately $\mathcal{O}(1)$ complexity, which is extremely helpful just as the backward propagation algorithm used in classical machine learning if a large number of parameters are involved.
This classical algorithm as illustrated in Algorithm~\ref{algo:rev-mode-grad} is termed as reverse-mode differentiation\cite{jones2020efficient, Guo_GITHUB_VQC_2020}.

\begin{algorithm}
    \caption{Reverse-mode algorithm to calculate $g_{i}=\frac{\partial E}{\partial \theta_{i}}$.}
        \SetAlgoLined
        \KwData{$\hat{H}$, $\{U_{0}(\theta_{0}), U_{0}(\theta_{0}), \ldots\}$, $|\Psi\rangle$}
        \KwResult{$g$: gradients of energy w.r.t. parameters $\{\theta_{0}, \theta_{1}, \ldots\}$}
        $N_{p}\leftarrow$ number of parameters, $g\leftarrow$empty array of length $N_{p}$\;
        \For{i=0; i$\le N_{p}-1$; i+=1}
        {
            $|\Psi\rangle\leftarrow U_{i}(\theta_{i})|\Psi\rangle$;
        }
        $|\Psi_{l} \rangle \leftarrow \hat{H} | \Psi \rangle$, $|\Psi_{r} \rangle \leftarrow |\Psi\rangle$\;
        \For{i=$N_{p}-1$; i$\ge 0$; i-=1}
        {
            $|\Psi_{r} \rangle \leftarrow U^{\dagger}_{i}(\theta_{i}) | \Psi_{r} \rangle$\;
            $g[i]$ = $2\times \text{Re}(\langle \Psi_{l} | \frac{\partial U_{i}(\theta_{i}) }{\partial \theta_{i}} \Psi_{r} \rangle$)\;
            $|\Psi_{l} \rangle \leftarrow U^{\dagger}_{i}(\theta_{i}) | \Psi_{l} \rangle$;
        }
    \label{algo:rev-mode-grad}
\end{algorithm}

{\system} implements reverse-mode differentiation for the brute-force backends. It should be noted that the reverse-mode differentiation algorithm is invalid on real quantum devices even if the original equation is used:
\begin{equation*}
    \frac{\partial E}{\partial \theta_{i}} = 2\times \text{Re}(\langle \Psi | U_{0}^{\dagger} \cdots U_{N_{p}-1}^{\dagger} \hat{H} U_{N_{p}-1} \cdots U_{i+1} \frac{\partial U_{i}}{\partial \theta_{i}} U_{i-1} \cdots U_{0} | \Psi \rangle),
\end{equation*}
since the derivative gates $\{\frac{\partial U_{i}(\theta_{i})}{\partial \theta_{i}} \}$ are generally non-unitary.
At the same time, using reverse-mode differentiation with the MPS simulator requires careful modifications to the algorithm and specially designed techniques due to the SVD truncation during the simulation. A naive implementation  of Algorithm~\ref{algo:rev-mode-grad} probably leads to suboptimal performance and numerical errors.

\section{Quantum algorithms}\label{sec:qalgo}
In the current release, {\system} provides a couple of VQE-based algorithms. The module \texttt{q2chem.qalgo} adopted several variational wave function ansatzes which can be used individually or collectively to solve for the eigenstates of the given chemical system. New quantum algorithms can also be implemented by users conveniently.

\subsection{Variational quantum circuit ansatz}
Introducing a parametric wave function $| \Psi(\theta) \rangle$, the lowest eigenvalue $E_{0}$ can be obtained variationally:
\begin{equation}
    E_{0} = \text{min}_{\theta} {\langle \Psi(\theta) | \hat{H} | \Psi(\theta) \rangle}.
\end{equation}
Such a protocol is implemented in the \texttt{q2chem.qalgo} module to obtain the eigenstates of a given Hamiltonian. Properly constructing a parametric quantum circuit, the wave function ansatz $| \Psi(\theta) \rangle$ is encoded into the quantum state of qubits. Combining the measurements for expectation value evaluation on a quantum computer and a numerical optimization algorithm on a classical computer, the variational procedure can then be performed in a hybrid quantum-classical way.

Generally, there are two broad types of variational quantum circuit ansatzes.\cite{CaoRomOls19} They are physically motivated ansatz (PMA) which is inspired by classical wave function methods that systematically approaches the exact electronic wave function and hardware heuristic ansatz (HHA) which considers specific hardware structure and employs entangling blocks. Both types are currently integrated into the \texttt{q2chem.qalgo} module. 

\begin{figure}
    \centering
    \subfigure[]
    {
        \includegraphics[width=0.45\textwidth]{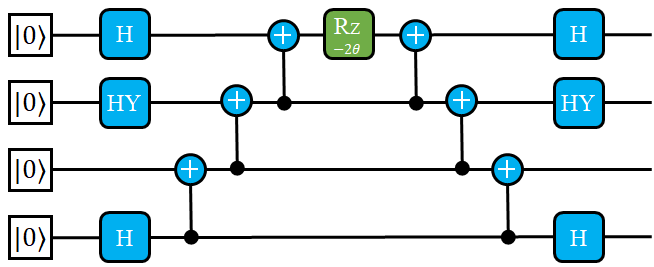}
    }
    \subfigure[]
    {
        \includegraphics[width=0.45\textwidth]{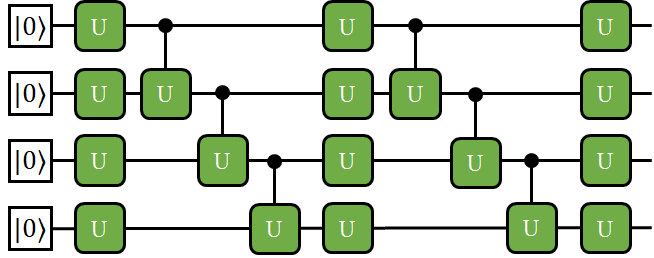}
    }
    \caption{(a) The quantum circuit corresponding to the operator $\exp{(i\theta \hat{\sigma^{x}} \hat{\sigma^{y}} \hat{\sigma^{z}} \hat{\sigma^{x}} )}$ and (b) the two-layer Kandala-Mezzacapo circuit which entangles all neighbouring qubits using the controlled-$U$ gate. Blue squares represent non-parametric gates while green represent parametric quantum gates such as $Rz$ and the three-parameter (controlled-)$U$ gate.}
    \label{fig:circ-example}
\end{figure}

Unitary coupled-cluster is one of the most commonly PMA for quantum computing. Generally, the UCC wave function is defined as
\begin{equation}
    \label{eq:ucc}
    | \Psi(\theta)^{UCC} \rangle = \exp{(\hat{T}(\theta) - \hat{T}^{\dagger}(\theta))} | \Psi^{HF} \rangle.
\end{equation}
In {\system}, the spin-adapt CCD0 cluster operators\cite{Scuseria_CCD0_1988, Bulik_CCD0_2015, Sokolov_CCD0_2020} are used to construct the UCCSD and UCCGSD wave functions. A fermion-to-qubit mapping is performed and first-order Trotter-Suzuki decomposition\cite{GriClaEco20, BabMcCWec15} is implemented to convert Equation~\ref{eq:ucc} into the product of Pauli strings:
\begin{equation}
    | \Psi(\theta)^{UCC} \rangle = \prod_{i} { \prod_{j} {\exp{(i\theta_{i} P_{ij})}} } | \Psi^{HF} \rangle,
\end{equation}
where the $i$-th fermion excitation operator $\hat{T}_i$ is transformed into the qubit form $\sum_{j} {P_{ij}}$. Each exponential term is mapped to quantum circuit following Algorithm~\ref{algo:exp-to-circ}. An example of the mapped circuit is present in Figure~\ref{fig:circ-example}a.

\begin{algorithm}
\caption{Map $\exp{(i\theta \hat{P})}$ to circuit. $HY$ is the Hadamard-Y gate defined as $HY = {\sqrt{2}}/{2}\times(Z + Y)$}
    \SetAlgoLined
    \KwData{$\hat{P}$, $\theta$}
    \KwResult{C: the quantum circuit}
    $N_{q}$ $\leftarrow$ number of qubits, C$\leftarrow$empty circuit\;
    \For{i=0; i$\le N_{q}-1$; i+=1}
    {
        $p_i$ = $P$[i]\;
        \uIf { $p_i$==$\hat{\sigma}_{x}$ }
        {
            C += $H_{i}$
        }
        \ElseIf { $p_i$==$\hat{\sigma}_{y}$ }
        {
            C += $HY_{i}$
        }
    }
    
    \For{i=$N_{q}-2$; i$\ge 0$; i-=1}
    {
        C += $CNOT_{(i+1, i)}$
    }

    C += $RZ(-2\theta)_{N_{q} - 1}$    

    \For{i=0; i$\le N_{q}-2$; i+=1}
    {
        C += $CNOT_{i+1, i}$
    }

    \For{i=0; i$\le N_{q}-1$; i+=1}
    {
        $p_i$ = $P$[i]\;
        \uIf { $p_i$==$\hat{\sigma}_{x}$ }
        {
            C += $H_{i}$
        }
        \ElseIf { $p_i$==$\hat{\sigma}_{y}$ }
        {
            C += $HY_{i}$
        }
    }
\label{algo:exp-to-circ}
\end{algorithm}

The Kandala-Mezzacapo circuit\cite{Kandala2017} is implemented as an HHA ansatz. If the type of entanglement gates, the ordering of entanglement qubits and the number of layers are set, {\system} automatically generates the parametric hardware efficient circuit for simulation. Figure~\ref{fig:circ-example}(b) shows an example of a two-layer Kandala-Mezzacapo circuit with all neighbouring qubits entangled by the three-parameter controlled-$U$ gate.

Several techniques are implemented in {\system} to reduce the computational overhead, including qubit tapering for Hamiltonian\cite{Bravyi_QUBIT_TAPER_2017}, qubit excitation based (QEB) operator\cite{Yordanov_QEB_2021}, Pauli entangler\cite{Ryabinkin_QCC_2021, Ryabinkin_iQCC_2020, Ryabinkin_iQCC_2021}, symmetry-based operator selection\cite{cao_SYMM_2022} and the ADAPT-VQE algorithm\cite{GriEcoBar19}. {\system} supports combination of above methods, for example, using Pauli entanglers which entangle at most 4 qubits together with ADAPT-VQE leads to an iterative-qubit-coupled-cluster (iQCC)\cite{Ryabinkin_iQCC_2020} like algorithm.

In addition to ground state methods, \texttt{q2chem.qalgo} module offers a couple of post-VQE algorithms for calculating excited states, including the variational quantum deflation (VQD)\cite{vqe-excited-vqd}, quantum subspace expansion (QSE)\cite{Mcclean17qse} and equation-of-motion (EOM)\cite{eom-1, eom-2, eom-3, Pau_QEOM_2019, FAN_PBC_2021} theory. VQD consecutively constructs an effective Hamiltonian, the lowest eigenvalue of which corresponds to the 1st, 2nd, 3rd, $\ldots$ excited state energy:
\begin{equation}
    \hat{H}^{i}_{eff} = \hat{H}^{i-1}_{eff} + \alpha_{i} |\Psi^{i-1}\rangle \langle \Psi^{i-1} |,
\end{equation}
where $\hat{H}^{0}_{eff}$ is the original second-quantized electronic Hamiltonian. QSE and EOM use a set of fermion excitation operators or a state-transfer operator to construct and solve a generalized eigenvalue problem $M C = S C E$ by additional quantum measurements on the top of the ground state circuit.
These methods are implemented in {\system} for calculating electron excitations, ionization potentials and electron affinity energies for both molecules and periodic systems.

\subsection{Parallel evaluation of an expectation value}\label{sec:qalgo:para-eval}
\begin{figure}
    \centering
    \includegraphics[width=0.49\textwidth]{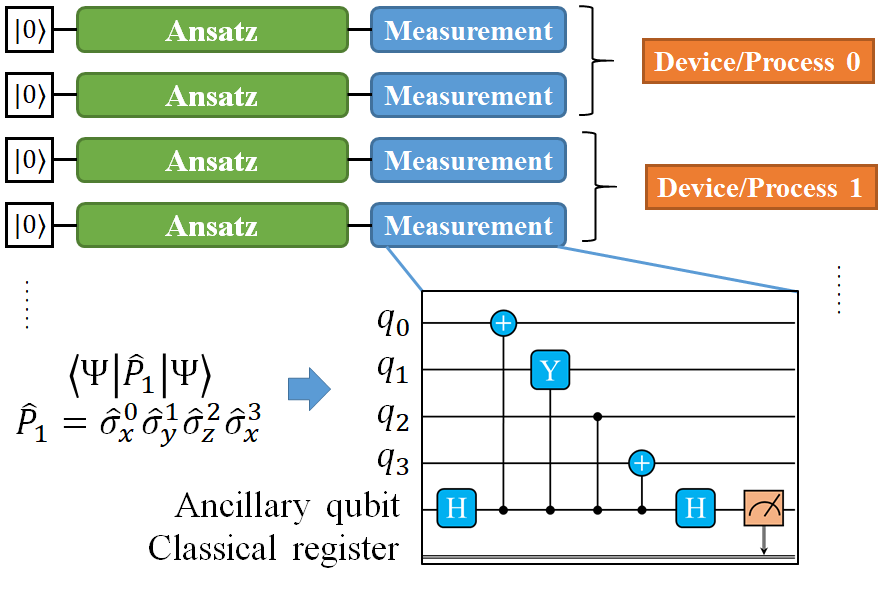}

    \caption{Evaluating the expectation of a linear combination of Pauli strings using multiple quantum devices or simulator processes. In this example, the \textit{measurement} parts are the Hadamard test circuits.}
    \label{fig:mol-circuit-para-meas}
\end{figure}
As shown in Equation~\ref{eq:qubit-ham-detail}, the Hamiltonian is expressed as the summation of a polynomial number of mutually uncorrelated Pauli strings. Expectation values of each Pauli string can thus be calculated independently. Figure~\ref{fig:mol-circuit-para-meas} gives an example of circuits used for evaluating expectation value $\sum_{i} { c_{i} \langle \Psi | {\hat{P}_{i}} | \Psi \rangle}$ for a given Hamiltonian containing a number of Pauli strings under some fermion-to-qubit transformation. For each of the circuits, the \textit{ansatz} parts are the same while the \textit{measurement} parts are constructed according to the specific form of Pauli string. During the calculation of expectation value, {\system} automatically distributes these circuits to different quantum devices or simulator processes. On each device, a subset of circuits are executed then measured. The measurement outcomes are then post-processed to calculate expectation values locally. Finally, the results are reduced across all the devices to obtain the total energy.

It should be noted that on real quantum devices, since the quantum states are non-replicable, all the circuits should be executed. However, on a classical simulator, the memory data of quantum states can be reused. Therefore, using a quantum circuit simulator, the \textit{ansatz} part only needs to be executed once, and the simulated quantum state can then be copied to each process for later measurements. This strategy is used in the simulations of Section~\ref{sec:app}.

\section{Applications}\label{sec:app}

We present several simulations to show the power of {\system}, including a scalability test for the MPS-based quantum circuit simulator and the numerical simulation results for ground- and excited-state calculations.

\subsection{Scalability benchmark}\label{sec:app:scala-bench}

\begin{figure}
    \centering

    \includegraphics[width=0.45\textwidth]{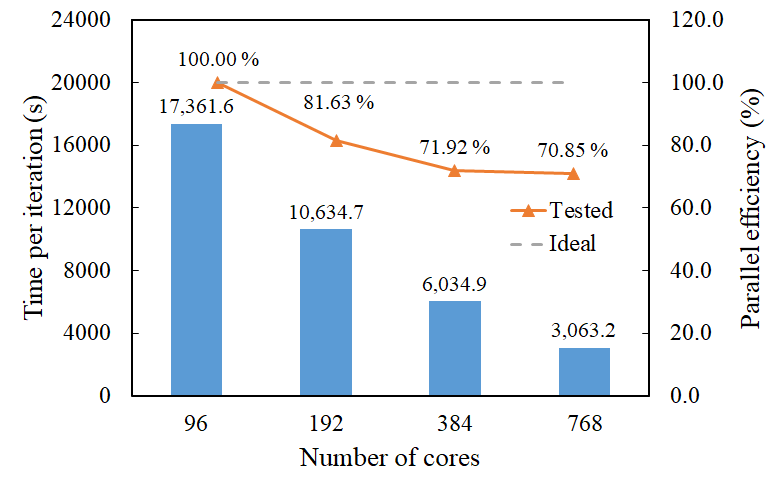}

    \caption{Simulating Cr$_2$ molecule using the MPS backend. STO-3G basis set and the symmetry-reduced UCCSD ansatz is used. The qubit Hamiltonian contains 305041 Pauli strings. 51 out of 3131 variational parameters (with random initial values) are selected, leading to a quantum circuit with 119884 gates. The upper bound for bond dimension is set to be 64. The time cost for one VQE iteration (including evolution of the quantum circuit and calculation of energy) is tested. The distributed parallelization is implemented using OpenMPI and Python's \texttt{mpi4py} package.}
    \label{fig:scaling-cr2}
\end{figure}
The parallel measurement for expectation value introduced in Section~\ref{sec:qalgo:para-eval} is extended to a two-level parallelism for the MPS simulator:
\begin{enumerate}
    \item The first level parallelization over Hamiltonian. Subsets of Pauli strings from the Hamiltonian in Equation~\ref{eq:qubit-ham} is distributed to each process. The expectation values of Pauli strings are calculated independently and a reduce-sum is performed across all processes to obtain the final energy.
    \item A second low-level parallelization using multi-thread parallelism on CPU or GPU, to accelerate the calculations of linear algebra routines such as matrix multiplication and singular value decomposition.
\end{enumerate}

The two-level strategy enables good parallel scalability if the adapted dynamical distribution algorithm is used to achieve load balance. To reduce memory usage, in the first level the quantum circuit is stored and evolved only on the $0$-th process. Although for small molecules such as H$_2$ (4 qubits), the circuit evolution may contribute over 80\% to the total execution time, the number of Pauli strings in the Hamiltonian will quickly go beyond $10^{5}$ due to the $\mathcal{O}(N^{4})$ scaling, and the time cost of circuit simulation will become negligible if larger systems with more that 12 qubits are involved. The scaling benchmark of the {\system} simulating a Cr$_2$ molecule (STO-3G basis set) using the MPS backend is given in Figure~\ref{fig:scaling-cr2}. {\system} achieves good parallel performance up to 768 CPU cores for this 72-qubit system.

\subsection{Numerical results}\label{sec:app:result}

\begin{figure}
    \centering

    \includegraphics[width=0.45\textwidth]{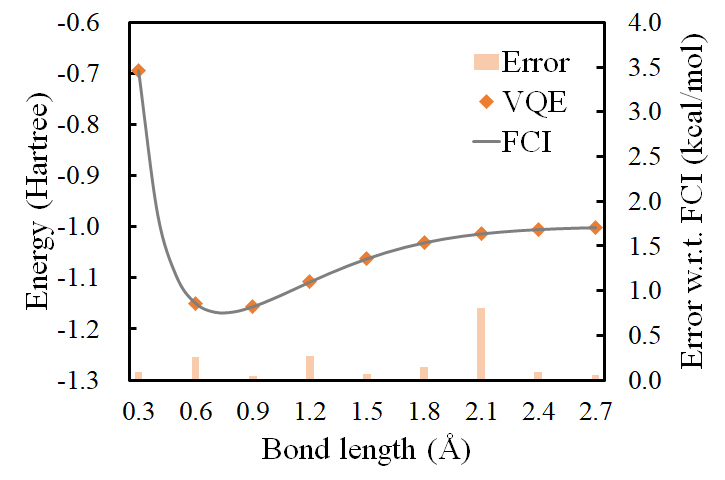}

    \caption{The VQE optimized potential energy curve of H$_2$ calculated by the MPS backend using ccj-pVDZ basis set. The FCI energies are obtained using the PySCF code, and the VQE results are calculated by interfacing with the external Julia-implemented MPS simulator.}
    \label{fig:h2-pes}
\end{figure}

Figure~\ref{fig:h2-pes} shows the potential energy curve of the H$_2$ molecule. The calculation is carried out by extending the simulators in \texttt{q2chem.qcirc} module to an external MPS circuit simulator written in Julia\cite{Guo_GITHUB_2020, Guo_QsimTN_2021}. Benefiting from the high-level modular structure introduced in Section~\ref{sec:design}, this external simulator and the parallelization techniques introduced in Section~\ref{sec:qalgo:para-eval} can be easily implemented within the framework of {\system} by adding a few lines of code. The ground-state energies are variationally optimized using ccj-pVDZ basis set\cite{Benedikt_CCJPVDZ_2008} and the symmetry-reduced UCCSD\cite{cao_SYMM_2022} circuit (leading to 40 qubits and 53 variational parameters). The BOBYQA optimizer is used to perform gradient-free optimizations. For each geometry, 2000 optimization steps are performed within 24 hours using 560 CPU cores.

\begin{figure}
    \centering

    \includegraphics[width=0.45\textwidth]{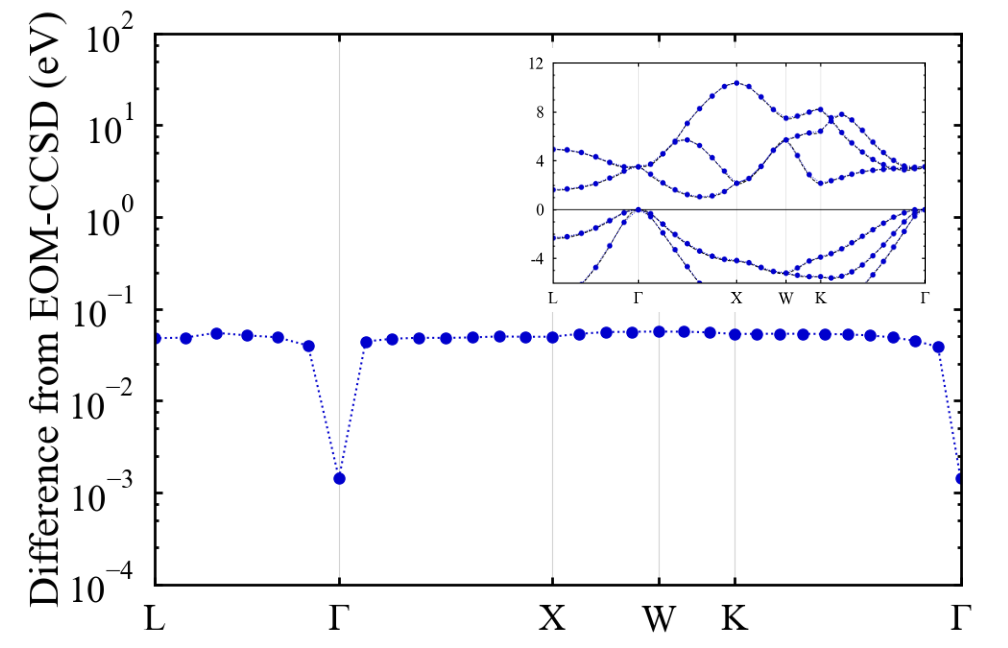}

    \caption{Energy difference between EOM-ADAPT-C and classical EOM-CCSD band structures for Si. Inset gives the EOM-ADAPT-C band structure calculated using {\system}. }
    \label{fig:si-band}
\end{figure}

Figure~\ref{fig:si-band} calculates quasi-particle band structures for silicon. For such periodic systems, the Hamiltonian and UCC-based wave function ansatz need to include the constraint of crystal momentum conservation\cite{Liu_PBC_2020}, which is automatically handled by the \texttt{q2chem.qchem} module. The simulation uses GTH-SVZ basis set with GTH-PADE pseudopotential. A UCCGSD operator pool with complementary operators \cite{FAN_PBC_2021} is employed together with the ADAPT algorithm for ground state ADAPT-C wave function. With a $1 \times 1 \times 1$ k-point grid, 16 qubits are simulated using the state vector backend. The EOM-ADAPT-C method \cite{FAN_PBC_2021} is used for band structure calculation, which achieves a mean absolute difference of 0.047 eV from EOM-CCSD and the deviation is as small as $\sim 10^{-3}$ eV at $\Gamma$ point (Figure~\ref{fig:si-band}). 

\section{Conclusion}
In this study, we demonstrate that the {\system} package is suitable for simulating and developing quantum algorithms for quantum chemistry applications. {\system} provides versatile functionalities for simulating ground- and excited-state properties. The simulator backend, including the parallelized MPS algorithm, achieves high performance for large-scale simulations up to 72 qubits using a moderate amount of computational resources. Directions for future development include more classical simulation methods including high-dimensional tensor network based methods, more integrated quantum algorithms, and high-performance quantum circuit optimization algorithms. With the flexibility to link to different quantum devices, {\system} can be used as a useful platform in pursuing practical quantum advantage.

\begin{acknowledgement}
		This work was partially supported by the NSFC (21825302), by the Fundamental Research Funds for the Central Universities (WK2060000018), by National Supercomputing Center in Jinan, and by the USTC Supercomputing Center.
\end{acknowledgement}






\bibliography{apssamp}

\end{document}